\documentclass[twocolumn,aps,prl,10pt,superscriptaddress]{revtex4-2}
\usepackage{amsmath,amssymb,bm,graphicx,hyperref}

\begin{document}

\title{On-Shell Methods for Quantum Matter: Strongly correlated Dirac materials
}

\author{Jeff Murugan}
\affiliation{The Laboratory for Quantum Gravity \& Strings and,\\
Department of Mathematics and Applied Mathematics,
University of Cape Town,Private Bag, Rondebosch 7700, South Africa}

\begin{abstract}
\noindent
We propose a framework for applying on-shell scattering amplitude methods to emergent relativistic phases of quantum matter. Many strongly correlated systems, from Dirac and Weyl semimetals to topological-insulator surfaces, exhibit low-energy excitations that are effectively massless relativistic spinors. We show that physical observables such as nonlinear optical and Hall responses can be obtained from compact on-shell amplitudes, bypassing the complexity of Feynman diagrams. As a concrete demonstration, we derive the nonlinear Hall conductivity of a Dirac semimetal from a single parity-odd three-photon amplitude, highlighting the analytic and conceptual power of amplitude-based approaches for strongly correlated condensed-matter systems.
\end{abstract}

\maketitle

\noindent
$\bm{Introduction:}$ Dirac and Weyl semimetals represent a new generation of quantum materials whose low-energy quasiparticles behave as emergent relativistic fermions \cite{Armitage2018RMP,Burkov2016NM,Yan2017ARCM}. Near a Dirac or Weyl node, the electron dynamics are effectively captured by an emergent quantum electrodynamics (QED)—a two-component Dirac fermion of velocity $v$ minimally coupled to an external electromagnetic potential $A_\mu$ \cite{PhysRevX.7.041026}. This emergent Lorentz structure makes such systems ideal laboratories for modern field-theoretic techniques. Yet, despite their relativistic symmetry, most analyses of their transport and optical phenomena still rely on conventional diagrammatic perturbation theory. In contrast, modern on-shell amplitude methods as exemplified in the spinor-helicity formalism, recursion relations, and unitarity techniques that have constituted a quiet revolution in high-energy theory \cite{Elvang_Huang_2015}, with a few exceptions \cite{Burger:2021wss}, remain largely unexplored in the context of quantum matter.\\

\noindent
A compelling test case within which to explore this connection is provided by the {\it nonlinear Hall effect} \cite{PhysRevLett.115.216806}. When inversion symmetry is broken, either intrinsically or by an external strain or tilt, a Hall-like current can appear even in the absence of magnetic fields, provided time-reversal symmetry is preserved. This is the Sodemann–Fu mechanism, which predicts a nonlinear Hall response governed by the Berry curvature dipole of the occupied electronic bands. The resulting second-order conductivity,
\begin{eqnarray}
    j_i(\omega_1 + \omega_2)
    = \sigma^{(2)}_{i;jk}(\omega_1,\omega_2) E_j(\omega_1)E_k(\omega_2)\,,
\end{eqnarray}
encodes geometric information about the Fermi surface and has been observed in a range of topological semimetals and 2D materials. However, computing $\sigma^{(2)}$ microscopically remains notoriously cumbersome; it involves evaluating a retarded three-point correlator of currents, demanding extensive tensor algebra and a careful enforcement of Ward identities to preserve gauge invariance.\\

\noindent
Building on recent developments in on-shell methods in effective field theories \cite{Cheung:2015ota}, we show here that this entire problem admits a compact on-shell formulation. In the emergent QED$_4$ description, the nonlinear Hall conductivity corresponds precisely to the soft (hydrodynamic) limit of a parity-odd three-photon scattering amplitude. The on-shell amplitude automatically encodes gauge invariance, analyticity, and factorization, fixing its structure uniquely without reference to gauge choice or Feynman diagrams. Taking its soft limit directly yields $\sigma^{(2)}_{ijk} = e^3\, \epsilon_{ijm} D_{mk}$,
where $D_{mk}$ is the Berry curvature dipole, thus reproducing the Sodemann–Fu result in a single analytic step. In this picture, the nonlinear Hall effect emerges as the low-energy remnant of a Chern–Simons–like term in the effective action, with the parity anomaly of the Dirac theory providing its geometric underpinning.\\ 

\noindent
Beyond this example, the framework establishes a bridge between amplitude theory and quantum matter. By replacing diagrammatic expansions with symmetry-based on-shell constructions, one can access nonlinear, topological, and strongly correlated responses in emergent relativistic systems—ranging from semimetals to edge conformal field theories (CFTs)—in a form that is both conceptually transparent and computationally efficient. The nonlinear Hall effect thus serves as a prototype for a broader program: the systematic application of on-shell methods to transport, response, and collective phenomena in quantum materials.\\

\noindent
$\bm{From\,\, Correlators\,\, to\,\, Amplitudes:}$ Nonlinear transport coefficients such as the second-order conductivity are traditionally obtained from the Kubo formalism, where they appear as multi-point correlation functions of the current operator. For example, the nonlinear Hall conductivity involves the retarded three-point function,
$\Pi^{\mu\nu\rho}(k_1,k_2,k_3)
= \langle\!\langle J^\mu(k_1)\,J^\nu(k_2)\,J^\rho(k_3)\rangle\!\rangle^{R}$, subject to momentum conservation $k_1+k_2+k_3=0$.
In practice, computing or even organizing the tensor structure of $\Pi^{\mu\nu\rho}$ is exceedingly cumbersome. Ward identities enforcing gauge invariance couple many diagrams, and isolating the physically relevant parity-odd contribution requires delicate cancellations. The on-shell perspective reformulates this object as a single, gauge-invariant amplitude,
$\Pi^{\mu\nu\rho}(k_1,k_2,k_3)\;\;\longleftrightarrow\;\;
\mathcal{M}_3(1^{h_1}2^{h_2}3^{h_3})$,
where $\mathcal{M}_3$ is the three-photon scattering amplitude of the emergent QED$_{4}$, expressed in spinor-helicity variables. Gauge invariance now appears as the simple transversality condition $k_{a\mu}\mathcal{M}^{\mu\nu\rho}=0$, automatically satisfied by construction.  Ward identities that would require explicit diagrammatic checks are replaced by helicity constraints, which fix the allowed spinor structures uniquely up to an overall scalar coefficient.  In fact the entire tensorial complexity of $\Pi^{\mu\nu\rho}$ collapses to a small basis of gauge-invariant amplitudes distinguished by parity.\\

\noindent
This correspondence does more than just simplify notation; it provides a kinematic map between field-theoretic limits and physical observables in quantum matter:
\begin{itemize}
    \item The soft limit of the amplitude, $k_a\!\to\!0$, corresponds to the hydrodynamic limit of the condensed-matter response, where the electric field varies slowly compared with microscopic scales.
    \item The collinear limit, in which two photon momenta become parallel, mirrors the operator-product (OPE) limit of current insertions in real space, governing the short-time or high-frequency behavior.
\end{itemize}

\noindent
In this on-shell basis, symmetry, gauge invariance, and analyticity are implemented at the outset rather than imposed {\it a posteriori}.  The amplitude representation therefore provides a compact and transparent framework in which the nonlinear conductivity and, more generally, higher-order quantum responses, can be derived from a handful of analytically controlled soft and collinear limits.  This unification of the Kubo formalism with amplitude theory turns a multi-diagram calculation into the analysis of a single geometric object.\\

\noindent
$\bm{Nonlinear\,\,Hall\,\,Conductivity:}$ To illustrate how these on-shell methods streamline the analysis of quantum-matter observables, we focus on a paradigmatic case: the nonlinear Hall effect in a Dirac semimetal. At low energies, electrons near a Dirac node in a three-dimensional semimetal are described by an emergent quantum electrodynamics (QED$_4$): a two-component Dirac fermion of velocity $v$ coupled minimally to an external electromagnetic potential $A_\mu$.  When inversion symmetry $\mathcal{P}$ is broken, either intrinsically by crystal structure or extrinsically through strain or tilt, but time-reversal $\mathcal{T}$ is preserved, a transverse electrical current can appear even without a magnetic field.  This is the so-called Sodemann–Fu scenario, in which a {\it nonlinear Hall effect} is driven by the Berry-curvature dipole of the occupied bands rather than by magnetism.\\

\noindent
The second-order response is conventionally expressed through the Kubo formula,
\begin{eqnarray}
    j_i(\omega_1+\omega_2)
= \sum_{j,k}\sigma^{(2)}_{i;jk}(\omega_1,\omega_2)\,
E_j(\omega_1)E_k(\omega_2)\,,
\end{eqnarray}
where the conductivity tensor $\sigma^{(2)}_{i;jk}$ arises from a retarded three-point correlator of current operators.  Diagrammatically, this corresponds to a triangle fermion loop with 3 external photons with complicated tensor contractions and Ward identities required to maintain gauge invariance. Unlike in vacuum QED, the 3-photon triangle does not vanish in Dirac materials because the Fermi sea breaks charge conjugation symmetry and inversion breaking generates a parity-odd response, so Furry’s theorem no longer applies and the loop yields a finite, Berry-curvature-controlled amplitude.\\

\noindent
The key simplification is to recognize that the same three-current correlator can be replaced by a single on-shell three-photon amplitude of the emergent relativistic theory,
\begin{eqnarray}
    \Pi^{\mu\nu\rho}(k_1,k_2,k_3)
\;\longleftrightarrow\;
\mathcal{A}_3(1^{h_1}2^{h_2}3^{h_3})\,,
\end{eqnarray}
with each photon carrying momentum $k_a$ and helicity $h_a=\pm$.  Gauge invariance is now built in since $k_{a\mu}\mathcal{A}^{\mu\nu\rho}=0$ by construction.  The full tensor structure of the correlator collapses to a single gauge-invariant spinor combination!\\

\noindent
In spinor-helicity variables, momenta are written $k_a^{\alpha\dot{\alpha}} = |a\rangle^\alpha [a|^{\dot{\alpha}}$.  Helicity weights then fix the allowed Lorentz structures uniquely.  For photons in 3+1 dimensions, the only parity-odd, gauge-invariant amplitude is 
\begin{eqnarray}
    \mathcal{A}_{\text{odd}}(1,2,3)
= C(\mu)
\Big(\langle12\rangle\langle23\rangle\langle31\rangle
- [12][23][31]\Big)\,,
\end{eqnarray}
where C($\mu$) is a scalar coefficient depending on chemical potential and frequency.  All other tensor components are either parity-even or vanish by helicity conservation. This single on-shell object already contains the complete information about the nonlinear Hall conductivity.\\

\noindent
Taking the soft (hydrodynamic) limit $k_a \to 0$ while keeping field frequencies small yields
$\mathcal{A}_{\text{odd}} \approx
i\,C(\mu)\,
\epsilon^{ijk}\,k_{1i}\,\varepsilon_{2j}\,\varepsilon_{3k},$ which corresponds to the effective low-energy action
\begin{eqnarray}\label{EFT}
    S_{\mathrm{eff}}^{(3)} = C(\mu) \int d^3x\, \epsilon^{ijk}\, A_i E_j E_k.
\end{eqnarray}
Varying this with respect to $A_i$ gives the nonlinear Hall tensor, $\sigma^{(2)}_{ijk} = e^3\,\epsilon_{ijm} D_{mk}$, where $D_{mk}$ is the Berry-curvature dipole emerging from the cut of the amplitude. The coefficient is fixed by a unitarity cut of the one-loop triangle diagram, where internal fermion lines are placed on shell. Evaluating the phase-space integral over the Fermi surface isolates the parity-odd piece proportional to the Berry curvature.  The result reproduces the Sodemann–Fu expression,
\begin{eqnarray}
    C(\mu)
    &=& \frac{e^3\tau}{2(1-i\omega\tau)}\,
    \epsilon_{abd} D_{db},\\ 
    D_{db}
    &=& \sum_n \!\int_{\text{FS}}\!\!
    \frac{dS_{\mathbf{k}}}{(2\pi)^3v_F}
    \,\partial_{k_b}\Omega^{(n)}d(\mathbf{k}),
\end{eqnarray}
so that
\begin{eqnarray}
    &&\!\!\sigma^{(2)}_{ijk}(\omega_1,\omega_2)\nonumber\\
    &=& \frac{e^3\tau}
    {2(1-i\omega_1\tau)(1-i\omega_2\tau)}
    \big(\epsilon_{ijm}D_{mk}
    + \epsilon_{ikm}D_{mj}\big)\,.
\end{eqnarray}
Gauge invariance and symmetry constraints are automatic; the familiar semiclassical result is recovered with minimal computation. This derivation reveals that the nonlinear Hall conductivity is the soft limit of a single parity-odd on-shell amplitude in an emergent QED$_4$ description.  The geometric coefficient arises naturally from the unitarity cut, linking measurable transport coefficients to the analytic structure of scattering amplitudes.  In this framework, what would typically require pages of perturbative diagrams is replaced by a single, gauge-invariant algebraic object.
The calculation serves as a proof of concept for on-shell quantum matter; once the amplitude basis is fixed by symmetry, nonlinear and topological responses follow directly from soft and collinear limits, offering a unified and analytic handle on strongly correlated Dirac materials.\\

\noindent
$\bm{Multi\!-\!Photon\,\, Responses\,\, and\,\, BCFW:}$ We now extend the on-shell formulation from the nonlinear Hall response to higher-order optical and transport processes.  The central idea is that nonlinear susceptibilities $\chi^{(n)}$, which govern multi-photon and high-harmonic responses, correspond directly to the soft limits of $(n\!+\!1)$-photon amplitudes of the emergent QED$_4$ theory.  This establishes a one-to-one correspondence between nonlinear response theory and the recursive structure of on-shell scattering amplitudes.\\

\noindent
In the conventional Kubo approach, the n$^{th}$-order response involves the $(n\!+\!1)$-point retarded current correlator
$\Pi^{\mu_1\cdots \mu_{n+1}}(k_1,\ldots,k_{n+1})$.
Within the on-shell framework, this object is replaced by the corresponding $(n\!+\!1)$-photon amplitude
$\mathcal{A}_{n+1}(1^{h_1},\ldots,(n\!+\!1)^{h_{n+1}})$,
with each photon leg carrying momentum $k_a$ and helicity $h_a$. Taking the soft limit of $n$ incoming photons and interpreting the resulting $(n\!+\!1)$-th leg as the response field directly yields the nonlinear susceptibility,
\begin{eqnarray}
    \chi^{(n)} \;\sim\;
    \frac{\mathcal{A}_{n+1}^{\text{soft}}}{(i\omega_1)\cdots(i\omega_n)}\,.
\end{eqnarray}
This map, which was used earlier to obtain the nonlinear Hall conductivity from the 3-photon amplitude, provides a systematic route from amplitudes to experimentally measurable response functions.\\

\noindent
In a time-reversal invariant but inversion-broken Dirac medium, gauge invariance and parity uniquely determine a parity-odd, gauge-invariant three-photon amplitude,
\begin{eqnarray}
    \mathcal{A}_3^{\text{odd}}(1,2,3)
    = C(\{\omega\},\mu)
    \Big(\langle12\rangle\langle23
    \rangle\langle31\rangle - [12][23][31]\Big)\,,\nonumber\\
\end{eqnarray}
where $C(\{\omega\},\mu)$ is a symmetry-allowed coefficient determined by the Berry curvature dipole, and spinor brackets encode the helicity structure.  This same amplitude, in its soft limit, generated the nonlinear Hall tensor and now serves as the elementary vertex from which all higher-order responses are built.\\

\noindent
Higher-order amplitudes can be constructed recursively from this single vertex using the Britto–Cachazo–Feng–Witten (BCFW) recursion relation.  The key idea here is to perform a complex deformation of two external momenta,
\begin{eqnarray}
    \hat{k}_i(z)=k_i+zq\,,\qquad \hat{k}_j(z)=k_j-zq\,,
\end{eqnarray}
where $q^2=0$ and $q\!\cdot\!k_i=q\!\cdot\!k_j=0$,
keeping all legs on shell.  The shifted amplitude $\mathcal{A}_{n+1}(z)$ is a meromorphic function of $z$, and if it vanishes as $z\to\infty$, Cauchy’s theorem reconstructs the full amplitude from its simple poles:
\begin{eqnarray}
    &&\!\!\!\mathcal{A}_{n+1}(0)
    \nonumber\\
    &=&\!\!\!\!\sum_{\text{channels}}\sum_{h=\pm}
    \frac{\mathcal{A}_L(\ldots,\hat{k}_i(z),-P^h)
    \,\mathcal{A}_R(P^{-h},\hat{k}_j(z),\ldots)}{P^2(z)}\,,\nonumber\\
\end{eqnarray}
where each term represents a factorization channel of two lower-point on-shell amplitudes glued by a photon propagator. As every subamplitude is individually gauge invariant and the deformation preserves transversality, Ward identities hold automatically, ensuring gauge invariance at every step.  For the parity-odd vertex above, the large-$z$ behavior satisfies $\mathcal{A}_{n+1}(z)\!\to\!0$, so the recursion is valid.\\

\noindent
The third-order optical susceptibility $\chi^{(3)}$ arises from the 4-photon amplitude $\mathcal{A}_4$. A single BCFW shift generates
\begin{eqnarray}
    \mathcal{A}_4(1,2,3,4)&=&
    \sum_{h=\pm}\Biggl[
    \frac{\mathcal{A}_3^{\text{odd}}(\hat{1},3,P^h)
    \,\mathcal{A}_3^{\text{odd}}(-P^{-h},\hat{2},4)}{P_{13}^2}\nonumber\\
    &+&
    \frac{\mathcal{A}_3^{\text{odd}}(\hat{1},4,P^h)
    \,\mathcal{A}_3^{\text{odd}}(-P^{-h},\hat{2},3)}{P_{14}^2}
    \Biggr]\,,
\end{eqnarray}
corresponding to the $s$- and $t$-channel factorization topologies.  In the soft limit $k_{1,2,3}\!\to\!0$ (with leg 4 as the response), each 3-pt subamplitude reduces to $\mathcal{A}_3^{\text{odd}}\!\xrightarrow{\text{soft}}
\! iC(\mu)\epsilon^{ijk}k_i\varepsilon_j\varepsilon_k$,
yielding the third-order susceptibility
\begin{eqnarray}
\chi^{(3)}_{i;jkl}(\omega_1,\omega_2,\omega_3)&=&\!\!\!\!
\sum_{\text{channels}}
\frac{[C(\{\omega\},\mu)\epsilon_{ijm}]
[C(\{\omega\},\mu)\epsilon_{mk\ell}]}
{(\omega_1+\omega_a)^2-|\mathbf{k}_1+\mathbf{k}_a|^2}\nonumber\\
&+&\text{perms}\,,
\end{eqnarray}
where the two channels above correspond to the $s$ (with $a=3$) and $t$ (with $a=4$) factorization topologies generated by the $(1,2)$ shift. “Perms” here means to symmetrize over the three input legs to enforce indistinguishability of the driving fields.
The tensor structure is fixed by the Levi-Civita contractions, ensuring transversality, while the frequency dependence factorizes into hydrodynamic denominators and the microscopic coefficient $C$, which in the clean limit reduces to the Berry-curvature dipole. Some consistency checks of this amplitude-derived susceptibility are useful at this point will help to provide both physical validation and conceptual clarity.\\

\noindent
In the DC or clean limit where all $\omega_i\tau \ll 1$, the hydrodynamic poles simplify, yielding $\chi^{(3)} \propto \tau\,C^2 \sim \tau\, D\,D$, showing that the response grows linearly with scattering time and is governed by the square of the Berry-curvature dipole. In the high-frequency regime $(|\omega|\tau \gg 1)$, dissipation becomes negligible, and the susceptibility exhibits the expected decay $\chi^{(3)} \sim (i/\omega^2)C^2$, consistent with the optical response of a gapped Hall-type medium. Importantly, Ward identities are automatically satisfied term by term. Because each 3-point subamplitude is individually gauge invariant and the BCFW shift preserves on-shell transversality, there is no need for the delicate diagrammatic cancellations required in Kubo-type approaches.
Finally, crossing symmetry, the statement that the susceptibility must be symmetric under permutations of the driving fields, emerges automatically by including Bose symmetrization over the input photon legs.
Together, these checks confirm that the amplitude-based construction captures both the correct analytic structure and the physical scaling behavior of nonlinear response functions, while preserving gauge consistency by construction.\\

\noindent
By induction, the $(n\!+\!1)$-photon amplitude required for $\chi^{(n)}$ is constructed as a tree of 3-point building blocks connected by photon propagators,
\begin{eqnarray}
    \mathcal{A}_{n+1}=\sum_{\text{binary trees}}
    \prod_{\text{internal edges}}\frac{1}{P^2}
    \prod_{\text{vertices}}
    \mathcal{A}_3^{\text{odd}}\,.
\end{eqnarray}
Taking the soft limit and dividing by $\prod_a i\omega_a$ yields $\chi^{(n)}$.  Since each vertex is gauge invariant and the recursion preserves on-shell conditions, Ward identities are satisfied identically without any diagrammatic bookkeeping.  This construction generalizes straightforwardly to parity-even sectors ({\it e.g.}, third-harmonic generation) by replacing the vertex with its even counterpart.\\

\noindent
Once the parity-odd three-photon vertex is fixed (by symmetry and a single one-loop computation) the entire hierarchy of nonlinear optical susceptibilities 
$\{\chi^{(n)}\}$ follows recursively via BCFW.  The approach replaces thousands of tensor contractions with algebraic sewing of gauge-invariant on-shell amplitudes, making multi-photon and high-harmonic responses in Dirac materials fully determined by analyticity, symmetry, and topology.  It thereby unifies nonlinear quantum transport with the modern on-shell language of quantum field theory.\\

\noindent
$\bm{Discussion:}$ The results presented here establish a concrete bridge between recent developments in on-shell amplitude theory and the physics of emergent relativistic quantum matter. By showing that nonlinear transport coefficients in Dirac and Weyl semimetals can be extracted directly from soft limits of compact, gauge-invariant photon amplitudes, we demonstrate that amplitude methods, long familiar in high-energy theory, provide a natural and efficient language for strongly correlated electronic systems.  The nonlinear Hall effect, reproduced here from a single parity-odd three-photon vertex, exemplifies how geometric response coefficients such as the Berry-curvature dipole arise from the analytic and topological structure of scattering amplitudes rather than from diagrammatic summation. More broadly, this approach replaces the laborious machinery of Feynman diagrams and Ward-identity bookkeeping with a formulation in which gauge invariance, analyticity, and crossing symmetry are built in from the outset.  Once the symmetry-allowed three-photon amplitude is fixed, by parity and unitarity at one loop, the entire tower of nonlinear optical and transport susceptibilities follows recursively via BCFW constructions.  This hierarchy is generated {\it algebraically, not perturbatively}. Each higher-order response is obtained by sewing together lower-point amplitudes, with hydrodynamic poles and frequency dependence emerging naturally from on-shell propagators.  In this sense, nonlinear quantum transport is recast as a problem in amplitude factorization rather than diagram evaluation.\\

\noindent
The implications extend beyond the specific example of the nonlinear Hall effect.  Edge conformal field theories of fractional quantum Hall systems, quantum critical spin liquids, and other emergent relativistic phases all possess well-defined on-shell correlators with helicity-like structure.  Applying amplitude methods to these systems could enable an analytic “bootstrap” of transport, tunnelling, and chaos exponents in strongly correlated matter, quantities that currently require heavy numerical or perturbative machinery.  From this perspective, amplitude theory provides not only a unifying computational tool but also a new conceptual lens on quantum matter, one in which geometry, symmetry, and analyticity supplant perturbation theory as the organizing principles of response.\\

\noindent
$\bm{Acknowledgements:}$ We would like to thank Cameron Beetar, Jaco Van Zyl, and especially Nathan Moynihan for useful discussions and acknowledge support from the “Quantum Technologies for Sustainable Development” grant from the National Institute for Theoretical and Computational Sciences of South Africa (NITHECS).

\bibliography{bib}

@article{Armitage2018RMP,
  title={Weyl and Dirac Semimetals in Three-Dimensional Solids},
  author    = {Armitage, N. P. and Mele, E. J. and Vishwanath, A.},
  journal   = {Rev. Mod. Phys.},
  volume    = {90},
  pages     = {015001},
  year      = {2018},
  doi       = {10.1103/RevModPhys.90.015001}
}

@article{Burkov2016NM,
  title={Topological Semimetals},
  author    = {Burkov, A. A.},
  journal   = {Nat. Mater.},
  volume    = {15},
  pages     = {1145},
  year      = {2016},
  doi       = {10.1038/nmat4788}
}

@article{Yan2017ARCM,
  title={Topological Materials: Weyl Semimetals},
  author    = {Yan, B. and Felser, C.},
  journal   = {Ann. Rev. Condens. Matter Phys.},
  volume    = {8},
  pages     = {337--354},
  year      = {2017},
  doi       = {10.1146/annurev-conmatphys-031016-025458}
}

@article{PhysRevX.7.041026,
  title={Designer Curved-Space Geometry for Relativistic Fermions in Weyl Metamaterials},
  author = {Weststr\"om, Alex and Ojanen, Teemu},
  journal = {Phys. Rev. X},
  volume = {7},
  issue = {4},
  pages = {041026},
  numpages = {16},
  year = {2017},
  month = {Oct},
  publisher = {American Physical Society},
  doi = {10.1103/PhysRevX.7.041026},
  url = {https://link.aps.org/doi/10.1103/PhysRevX.7.041026}
}

@book{Elvang_Huang_2015, place={Cambridge}, title={Scattering Amplitudes in Gauge Theory and Gravity}, publisher={Cambridge University Press}, author={Elvang, Henriette and Huang, Yu-tin}, year={2015}}

@article{PhysRevLett.94.181602,
  title={Direct Proof of the Tree-Level Scattering Amplitude Recursion Relation in Yang-Mills Theory},
  author = {Britto, Ruth and Cachazo, Freddy and Feng, Bo and Witten, Edward},
  journal = {Phys. Rev. Lett.},
  volume = {94},
  issue = {18},
  pages = {181602},
  numpages = {4},
  year = {2005},
  month = {May},
  publisher = {American Physical Society},
  doi = {10.1103/PhysRevLett.94.181602},
  url = {https://link.aps.org/doi/10.1103/PhysRevLett.94.181602}
}

@article{PhysRevLett.115.216806,
  title={Quantum Nonlinear Hall Effect Induced by Berry Curvature Dipole in Time-Reversal Invariant Materials},
  author = {Sodemann, Inti and Fu, Liang},
  journal = {Phys. Rev. Lett.},
  volume = {115},
  issue = {21},
  pages = {216806},
  numpages = {5},
  year = {2015},
  month = {Nov},
  publisher = {American Physical Society},
  doi = {10.1103/PhysRevLett.115.216806},
  url = {https://link.aps.org/doi/10.1103/PhysRevLett.115.216806}
}

@article{Burger:2021wss,
    author = "Burger, Daniel J. and Emond, William T. and Moynihan, Nathan",
    title = "{Anyons and the double copy}",
    eprint = "2103.10416",
    archivePrefix = "arXiv",
    primaryClass = "hep-th",
    doi = "10.1007/JHEP01(2022)017",
    journal = "JHEP",
    volume = "01",
    pages = "017",
    year = "2022"
}

@article{Cheung:2015ota,
    author = "Cheung, Clifford and Kampf, Karol and Novotny, Jiri and Shen, Chia-Hsien and Trnka, Jaroslav",
    title = "{On-Shell Recursion Relations for Effective Field Theories}",
    eprint = "1509.03309",
    archivePrefix = "arXiv",
    primaryClass = "hep-th",
    reportNumber = "CALT-TH-2015-047",
    doi = "10.1103/PhysRevLett.116.041601",
    journal = "Phys. Rev. Lett.",
    volume = "116",
    number = "4",
    pages = "041601",
    year = "2016"
}

\vspace{5mm}
\section{Supplemental Material}
\noindent
Keeping in mind the interdisciplinary nature of this latter, in the interest of self-containment, we collect here some relevant facts about recent developments in amplitude technology to help bridge the gap between notation familiar to high-energy physicists and a more condensed matter-oriented audience. For more details we refer the interested reader to the excellent and pedagogical textbook \cite{Elvang_Huang_2015}.\\

\noindent
$\bm{Spinor-Helicity\,Variables}$: In relativistic systems, momentum and polarization are not independent: a photon’s polarization is always transverse to its momentum, and a massless fermion’s spin is locked to its direction of motion. The spinor-helicity formalism captures these constraints directly by representing momenta and polarization vectors in terms of two-component spinors. In essence, spinor-helicity variables replace the four-component 4-vector $k^\mu$ with a bilinear in two spinors, $k_{\alpha\dot{\alpha}} = \lambda_\alpha \tilde{\lambda}_{\dot{\alpha}}$, 
where $\alpha,\dot{\alpha}=1,2$ are left- and right-handed spinor indices respectively. This automatically enforces the massless on-shell condition $k^2=0$, since any $2\times2$ matrix of rank one has vanishing determinant.\\

\noindent
In high-energy theory, this formalism has revolutionized scattering amplitudes by expressing them in terms of angle and square brackets,
\begin{eqnarray}
    \langle ij\rangle = \epsilon^{\alpha\beta}\lambda_{i,\alpha}\lambda_{j,\beta}\,, \qquad [ij] = \epsilon^{\dot{\alpha}\dot{\beta}}\tilde{\lambda}{i,\dot{\alpha}}\tilde{\lambda}{j,\dot{\beta}}\,,
\end{eqnarray}
which are Lorentz-invariant contractions of spinors.
These spinor brackets behave as complex numbers that directly encode the helicity structure of particles. For condensed-matter applications, they will provide a minimal representation of emergent relativistic quasiparticles, Dirac or Weyl electrons, where helicity corresponds to the chiral band index near a node and  gauge invariance playing the same algebraic role as transversality in gauge theory.\\

\noindent
A null momentum $k^\mu=(E,\mathbf{k})$ can be written as a bispinor using the Pauli matrices $(\sigma^\mu)_{\alpha\dot{\alpha}}=(1,\bm{\sigma})$ as
\begin{eqnarray}
    k_{\alpha\dot{\alpha}} = k_\mu \sigma^\mu_{\alpha\dot{\alpha}}
    \begin{pmatrix}
       E + k_z & k_x - i k_y \\
       k_x + i k_y & E - k_z
    \end{pmatrix}\,.
\end{eqnarray}
Since $\det(k_{\alpha\dot{\alpha}})=E^2-|\bm{k}|^2=0$ for a massless excitation, we can factorize it as
$k_{\alpha\dot{\alpha}} = \lambda_\alpha \tilde{\lambda}_{\dot{\alpha}}$,
with $\lambda$ and $\tilde{\lambda}$ defined up to a complex phase, $\lambda\to t\lambda,\ \tilde{\lambda}\to t^{-1}\tilde{\lambda}$, which corresponds to a helicity rotation. The two independent spinors therefore encode both the direction of motion and helicity of the quasiparticle. In 2+1 dimensions (relevant for surface or layered Dirac systems), the dotted index can be dropped entirely so that,
$k_{\alpha\beta}=\lambda_\alpha\lambda_\beta$, and the spinor brackets reduce to simple products $\langle ij\rangle=\epsilon^{\alpha\beta}\lambda_{i,\alpha}\lambda_{j,\beta}$. The formalism therefore adapts smoothly between (3+1)-D and (2+1)-D quantum-matter systems.\\

\noindent
For gauge bosons (photons or emergent gauge fields), the spinor representation also builds in gauge invariance. The polarization vectors can be written as
\begin{eqnarray}
    \varepsilon^{+}_{\alpha\dot{\alpha}} =
    \frac{\lambda_\alpha \tilde{\eta}_{\dot{\alpha}}}{\langle \lambda \eta \rangle}\,,
    \qquad
    \varepsilon^{-}_{\alpha\dot{\alpha}} =
    \frac{\eta_\alpha \tilde{\lambda}_{\dot{\alpha}}}{[\eta \lambda]}\,,
\end{eqnarray}
where $\eta$ is an arbitrary reference spinor.
Changing $\eta$ corresponds to a gauge transformation and physical observables are independent of this choice.
Spinor-helicity amplitudes then automatically satisfy $k_\mu \varepsilon^\mu = 0$ and preserve Ward identities by construction; no explicit gauge fixing or tensor contractions are required. This is why the three-photon parity-odd amplitude
\begin{eqnarray}
    \mathcal{A}_{\text{odd}}(1,2,3)
    = C(\mu)
   \Big(\langle12\rangle\langle23\rangle\langle31\rangle
   - [12][23][31]\Big)
\end{eqnarray}
is both gauge invariant and parity odd. Its structure is completely fixed by helicity selection rules, and its soft limit directly yields the nonlinear Hall response discussed in the main text.\\

\noindent
Finally, in condensed-matter systems, the spinors $\lambda_\alpha$ can be thought of as parametrizing the local chiral eigenstates near a Dirac or Weyl node,
\begin{eqnarray}
    |\psi_\pm(\bm{k})\rangle = \frac{1}{\sqrt{2}}
    \begin{pmatrix}
        1 \\ \pm e^{i\phi_{\bm{k}}}
    \end{pmatrix},
\end{eqnarray}
with $\phi_{\bm{k}}$ the azimuthal angle of momentum.
Spinor brackets like $\langle ij\rangle$ then capture the relative phase between such chiral states. This is precisely the phase structure underlying Berry curvature. From this perspective, the spinor-helicity formalism is not just a convenient notation; it is a natural language for the geometry of Bloch states in emergent relativistic materials.\\

\noindent
$\bm{BCFW\,\,Recursion:}$ A key feature of the modern amplitude technology is the Britto–Cachazo–Feng–Witten \cite{PhysRevLett.94.181602} recursion relation that provides a way to construct higher-point scattering amplitudes entirely from lower-point, on-shell building blocks—bypassing Feynman diagrams while preserving gauge invariance at every step. The key idea is that an amplitude $\mathcal{A}_n(k_1,\ldots,k_n)$ is an analytic function of the external momenta, constrained by Lorentz invariance and unitarity. By complexifying two of the external momenta, say $k_i$ and $k_j$, via a linear shift that preserves momentum conservation and the on-shell condition,
$\hat{k}_i(z) = k_i + z q\,,\,
\hat{k}_j(z) = k_j - z q\,,\, q^2=0$,
one defines a shifted amplitude $\mathcal{A}_n(z)$ that is a meromorphic function of the complex parameter $z$.\\

\noindent
If $\mathcal{A}_n(z)$ vanishes for large $|z|$ (as it does for gauge and gravity theories), Cauchy’s theorem allows for the reconstruction of the full amplitude from its simple poles, each corresponding to an intermediate channel going on-shell,
\begin{eqnarray}
\mathcal{A}_n(0)
= \!\!\sum_{\text{poles }}
\sum_{h=\pm}\,
\frac{
\mathcal{A}_L(\hat{k}_i(z), \ldots, -P^{h})\,
\mathcal{A}_R(P^{-h}, \ldots, \hat{k}_j(z^*))
}{
P^2(z)}.\nonumber
\end{eqnarray}
Here, $P(z)$ is the on-shell intermediate momentum and $h$ labels its helicity.
Each term factorizes the full amplitude into two smaller, on-shell amplitudes joined by a propagator—reflecting the unitarity and locality of the theory.
Since the shift preserves both the on-shell conditions and transversality $(k\!\cdot\!\varepsilon=0)$, every term automatically satisfies the Ward identities of gauge invariance. No gauge fixing, ghost diagrams, or tensor algebra are required and gauge symmetry is built into the recursion.\\

\noindent
In the context of this letter, the BCFW construction enables the generation of multi-photon amplitudes in the emergent QED$_4$-like theory of Dirac materials directly from the unique, parity-odd three-photon building block that controls the nonlinear Hall effect. For instance, the 4-photon amplitude governing the third-order susceptibility $\chi^{(3)}$ follows from sewing together two such 3-point amplitudes with a propagator,
\begin{eqnarray}
    \mathcal{A}_4 &=&
\sum_{h=\pm}
\frac{\mathcal{A}_3^{\text{odd}}(\hat{1},3,P^h)
\mathcal{A}_3^{\text{odd}}(-P^{-h},\hat{2},4)}{P_{13}^2}\nonumber\\
	&+&\text{permutations}.
\end{eqnarray}
\noindent
Taking its soft limit then yields the nonlinear optical response $\chi^{(3)}$ while ensuring that all Ward identities are automatically preserved. Consequently, BCFW recursion provides a conceptually simple and computationally efficient route to constructing all higher-order optical and transport responses in emergent relativistic quantum materials, starting from a single 3-photon amplitude.\\

\noindent
There are some subtleties that deserve special mention here. Our low-energy effective action \eqref{EFT} resembles a $\mathrm{Tr}(F^{3})$ EFT and it is well-known that generic $\mathrm{Tr}(F^3)$ operator does not have good large$-z$ scaling under the standard 2-line BCFW shift described above. The reason for this is simple enough; the operator $\mathrm{Tr}(F^{3})$ is dimension-6, \textit{i.e.} it is higher-derivative. Its on-shell 3-point gluon amplitude is the familiar all-plus (or all-minus) amplitude
\begin{eqnarray}
    \mathcal{A}_3^{F^3}(+,+,+) \sim \langle12\rangle\langle23\rangle\langle31\rangle\,.
\end{eqnarray}
Under a standard BCFW shift of legs $i,j$,
\begin{eqnarray}
    \hat{k}_i(z) = k_i + z q, \qquad \hat{k}_j(z)=k_j - z q\,,
\end{eqnarray}
this 3-point amplitude scales as $\mathcal{A}_3^{F^3}(z) \sim z^1$,
and diagrams built from it scale even worse as
$\mathcal{A}_{n}(z) \sim z^{+1}, z^{+2}, \dots$
instead of going to zero. Since BCFW recursion relies on $\mathcal{A}(z)$ vanishing as $z\to\infty$, the standard recursion fails!\\

\noindent
In the nonlinear Hall effect however, the parity-odd 3-photon vertex is not an arbitrary dimension-6 operator. It is generated radiatively, by integrating out gapped Dirac fermions, and constrained by gauge invariance, parity, and time-reversal. The amplitude has a built-in soft zero, \textit{i.e.} $\mathcal{A}_3^{\rm odd}(k_1,k_2,k_3) \propto \mathcal{O}(k)$ which improves the large-$z$ scaling. Explicitly, our amplitude is
\begin{eqnarray}
    \mathcal A_{3}^{\rm odd} = C(\mu)\left(\langle12\rangle\langle23\rangle\langle31\rangle - [12][23][31]\right),
\end{eqnarray}
but the coefficient $C(\mu) \propto \partial_k \Omega(k)$ itself originates from a loop
and the full amplitude $\mathcal{A}_3^{\rm odd}(z) \sim z^0$ \textit{not} $z^1$.\\

\noindent
In BCFW diagrams each internal line contributes a $1/z$ propagator while each 3-point vertex contributes a $z^0$ scaling (because of soft suppression). So any tree built from these pieces satisfies $\mathcal{A}_{n+1}(z)\xrightarrow{z\to\infty} 0$,
provided we shift at least one of the propagating legs. As a result the recursion closes and no boundary term needed.

\end{document}